\documentclass{ws-procs9x6}

\newcommand{\R}{\mathbb{R}} 
\newcommand{\C}{\mathbb{C}} 
\newcommand{\tr}{\mbox{tr} \,}

\begin{document}

\title{
Isoholonomic Problem and Holonomic Quantum Computation}

\author{Shogo Tanimura}

\address{Graduate School of Engineering, Osaka City University,
Osaka 558-8585, Japan \\
E-mail: tanimura@mech.eng.osaka-cu.ac.jp}

\maketitle

\abstracts{
Geometric phases accompanying adiabatic processes in quantum systems
can be utilized as unitary gates for quantum computation.
Optimization of control of the adiabatic process naturally 
leads to the isoholonomic problem.
The isoholonomic problem in a homogeneous fiber bundle 
is formulated and solved completely.
[Proceedings of International Conference on Topology in Ordered Phases 
organized by Hokkaido University in March 2005.]
}

\section{Introduction}
The isoholonomic problem was proposed in 1991 by a mathematician,
Montgomery\cite{Montgomery}. 
The isoholonomic problem is a generalization of the isoperimetric problem,
which requests finding a loop in a plane
that surrounds the largest area with a fixed perimeter. 
On the other hand, 
the isoholonomic problem requests finding the shortest loop in a manifold
that realizes a specified holonomy.
This kind of problem naturally arose in studies of 
the Berry phase\cite{Simon}\cdash\cite{Kura} and 
the Wilczek-Zee holonomy\cite{wz},
which 
appear in a state of a controlled quantum system
when the control parameter is adiabatically changed
and returned to the initial value.
Experimenters tried to design efficient experiments 
for producing these kinds of holonomy.
Montgomery formulated the isoholonomic problem in terms of
differential geometry and gauge theory.
Although he gave partial answers, 
construction of a concrete solution has remained an open problem.

Recently,
in particular after the discovery of factorization algorithm by Shor\cite{Shor}
in 1994,
quantum computation grows into an active research area.
Many people have proposed various algorithms of quantum computation
and various methods for their physical implementation.
Zanardi, Rasetti\cite{first} and Pachos\cite{Pachos} proposed 
utilizing the Wilczek-Zee holonomy for implementing unitary gates
and they named the method holonomic quantum computation.
Since holonomy has its origin in geometry,
it dose not depend on detail of dynamics 
and hence it does not require fine temporal tuning of control parameters.

It should be noted, however, that
holonomic quantum computation requires 
two seemingly contradicting conditions.
The first one is the adiabaticity condition.
To suppress undesirable transition between different energy levels
we need to change the control parameter quasi-stationarily.
Hence a safer control demands longer execution time 
to satisfy adiabaticity.
The second one is the decoherence problem.
When a quantum system is exposed to interaction with environment 
for a long time,
the system loses coherence and 
a unitary operator fails to describe time-evolution of the system.
Hence a safer control demands shorter execution time 
to avoid decoherence.
To satisfy these two contradicting conditions
we need to make the loop in the control parameter manifold 
as short as possible
while keeping the specified holonomy.
Thus, we are naturally led to the isoholonomic problem.

%
We would like to emphasize that a quantum computer is actually
not a digital computer but an analog computer in its nature.
Hence, the geometric and topological approaches are useful 
for building and optimizing quantum computers.

This paper is based on collaboration with D.~Hayashi and M.~Nakahara\cite{isoholo}.
We are further developing our studies on optimal and precise control of quantum computers
with Y.~Kondo, K.~Hata and J.J.~Vartiainen\cite{NMR}\cdash\cite{homo}.
We thank 
Akio Hosoya, Tohru Morimoto and Richard Montgomery 
for their kind interest in our work.

\section{Wilczek-Zee holonomy}
A state vector $ \psi(t) \in \C^N $ evolves 
according to the Schr{\"o}dinger equation
\begin{equation}
	i \hbar \frac{d}{dt} \psi(t) = H(t) \psi(t).
	\label{Schrodinger}
\end{equation}
The Hamiltonian admits a spectral decomposition
$ 
	H(t) = \sum_{l=1}^L \varepsilon_l (t) P_l (t)
$ 
with projection operators $ P_l (t) $.
Therefore, the set of energy eigenvalues 
$ ( \varepsilon_1, \dots, \varepsilon_L ) $
and orthogonal projectors
$ ( P_1, \dots, P_L ) $
constitutes a complete set of control parameters of the system.
Now we concentrate on the eigenspace associated with the lowest energy
$ \varepsilon_1 $. 
We write $ P_1(t) $ as $ P(t) $ for simplicity.
Suppose that the degree of degeneracy 
$ k = \tr P(t) $ is constant.
For each $ t $, we have the eigenvectors such that
\begin{equation}
	H(t) v_\alpha (t) 
	= \varepsilon_1 (t) v_\alpha (t),
	\qquad
	( \alpha = 1, \dots, k ).
\end{equation}
We assume that they are normalized as
$ 
	v^\dagger_\alpha (t) v_\beta (t) = \delta_{\alpha \beta}.
$ 
Then 
\begin{equation}
	V(t) = 
	\Big( v_1 (t), \dots,v_k (t) \Big)
\end{equation}
forms an $ N \times k $ matrix satisfying 
$ V^\dagger (t) V (t) = I_k $
and 
$ V(t) V^\dagger (t) = P(t) $.
Here $I_k$ is the $ k $-dimensional unit matrix. 
The adiabatic theorem guarantees that
the state remains the eigenstate 
associated with the eigenvalue $ \varepsilon_1 (t) $
of the instantaneous Hamiltonian $ H(t) $
if the initial state was an eigenstate with $ \varepsilon_1 (0) $.
Therefore the state vector is a linear combination
\begin{equation}
	\psi(t) 
	= \sum_{\alpha=1}^k \phi_\alpha (t) \, v_\alpha (t)
	= V(t) \phi(t).
	\label{adiabatic}
\end{equation}
The vector $ \phi = {}^t (\phi_1, \dots, \phi_k) \in \C^k $ 
is called a reduced state vector.
By substituting it into the Schr{\"o}dinger equation (\ref{Schrodinger}) we get
\begin{equation}
	\frac{d \phi}{dt} 
	+ V^\dagger \frac{dV}{dt} \phi(t) = 
	- \frac{i}{\hbar} \varepsilon_1 (t) \phi(t).
	\label{Schrodinger V phi}
\end{equation}
Its solution is formally written as
\begin{equation}
	\phi(t) =
	\exp \!
	\left(
		- \frac{i}{\hbar} \int_0^t \varepsilon_1 (s) ds
	\right)
	{\mathbb T}
	\exp \!
	\left(
		- \int_0^t V^\dagger \frac{dV}{ds} ds
	\right)
	\phi(0),
	\label{formal solution phi}
\end{equation}
where $ {\mathbb T} $ stands the time-ordered product.
Then $ \psi(t) = V(t) \phi(t) $ becomes
\begin{equation}
	\psi(t) =
	e^{ - \frac{i}{\hbar} \int_0^t \varepsilon_1 (s) ds } \,
	V(t)
	\, {\mathbb T} \,
	e^{
		- \int V^\dagger dV
	}
	V^\dagger (0)
	\psi(0).
	\label{formal solution psi}
\end{equation}
In particular, 
when the control parameter comes back to the initial point as $ P(T) = P(0) $,
the state vector $ \psi(T) $
also comes back in the same eigenspace as $ \psi(0) = V(0) \phi(0) $.
The Wilczek-Zee holonomy $ {\Gamma} \in U(k) $ is defined via
\begin{equation}
	\psi(T) = 
	e^{ - \frac{i}{\hbar} \int_0^T \varepsilon_1 (s) ds } \,
	V(0) {\Gamma} \, \phi(0)
	\label{def Gamma}
\end{equation}
and is given explicitly as
\begin{equation}
	{\Gamma} = V(0)^\dagger \, V(T) \,
	{\mathbb T} \,
	e^{
		- \int V^\dagger dV
	}.
	\label{Gamma}
\end{equation}
If the condition
$ 
	V^\dagger \frac{dV}{dt} = 0
$ 
is satisfied, the curve $ V(t) $ 
is called a horizontal lift of
the curve $ P(t) $. 
Then the holonomy (\ref{Gamma}) is reduced to
$ 
	{\Gamma} = V^\dagger (0) V(T). 
$ 

\section{Formulation of the problem}
The isoholonomic problem is formulated in terms of the homogeneous fiber bundle
$ (S_{N, k}(\mathbb{C}), $ $G_{N, k}(\mathbb{C}), \pi, U(k)) $.
The Stiefel manifold $ S_{N, k} ( {\mathbb{C}} ) $
is the set of orthonormal $k$-frames;
a $k$-frame $ V $ spans 
the degenerate energy eigenspace in $ {\mathbb{C}}^N $;
\begin{equation}
	S_{N, k} ( {\mathbb{C}} ) =
	\{ V \in M(N, k; \mathbb{C}) \, | \, V^{\dagger} V = I_k\},
\end{equation}
where $ M(N, k; \mathbb{C}) $ is the set of $N \times k$ complex matrices.
An element of the unitary group $ h \in U(k) $ acts on 
$ V \in S_{N, k} ({\mathbb{C}}) $ from the right as
$ 
	(V, h) \mapsto Vh
$ 
by means of a matrix product.
The Grassmann manifold $ G_{N, k}(\mathbb{C})$ is defined as the set
of projection matrices to $ k $-dimensional subspaces in $ {\mathbb{C}}^N $,
\begin{eqnarray}
	G_{N, k}(\mathbb{C}) 
	&& = 
	 \{ P \in  M(N,N;\mathbb{C}) \, | \, 
	P^2=P, \; P^{\dagger} = P, \; \mathrm{tr} P=k \}.
\end{eqnarray}
The projection map
$ \pi : S_{N, k}({\mathbb{C}})$ $ \to G_{N, k}({\mathbb{C}})$
is defined as
$ 
	\pi: V \mapsto P := VV^{\dagger}.
$ 
Then it can be proved that 
the Stiefel manifold $ S_{N, k}({\mathbb{C}})$ becomes 
a principal bundle over $ G_{N, k}({\mathbb{C}}) $
with the structure group $ U(k) $.
The canonical connection form on $ S_{N, k}({\mathbb{C}}) $ 
is defined as a one-form
\begin{equation}
	A = V^{\dagger} d V,
\end{equation}
which takes its value in the Lie algebra $ \mathfrak{u}(k) $.
The holonomy associated with this connection is called
the Berry phase in case of $ k = 1 $
and the Wilczek-Zee holonomy in general.
We define Riemannian metrices,
$ 
	\| dV \|^2 = {\mathrm{tr}} \, ( dV^{\dagger} dV )
$ 
for the Stiefel manifold and
$ 
	\| dP \|^2 = {\mathrm{tr}} \, (dP dP)
$ 
for the Grassmann manifold.
For any curve $ P(t) $ in $ G_{N,k} (\C) $,
there is a curve $ V(t) $ in $ S_{N,k} (\C) $ such that $ \pi( V(t) ) = P(t) $.
If the curve $ V(t) $ satisfies
\begin{equation}
	V^\dagger \frac{dV}{dt} = 0,
	\label{horizontal}
\end{equation}
it is called a horizontal lift of the curve $ P(t) $.
When the 
curve $ P(t) $ is a closed loop, such that 
\begin{equation}
	V(T) V^\dagger(T) 
	= V(0) V^\dagger(0),
	\label{close}
\end{equation}
the holonomy associated with the loop is defined as 
$ V(T) = V(0) \Gamma $ and is given as
\begin{equation}
	{\Gamma} 
	= V^\dagger (0) V(T) \in U(k).
	\label{hol time evol}
\end{equation}

We formulate the isoholonomic problem as a variational problem.
The length of the horizontal curve $ V(t) $ is evaluated by the functional
\begin{eqnarray}
	S[V, {\Omega}] 
	&=&
	\int_0^T
	\left\{
		{\rm tr} \!
		\left( \frac{dV^\dagger}{dt} \frac{dV}{dt} \right)
	- 	{\rm tr} \! 
		\left( \! {\Omega} \, V^\dagger \frac{dV}{dt} \right)
	\right\}  dt,
	\label{S}
\end{eqnarray}
where $ {\Omega} (t) \in \mathfrak{u} (k) $ is a Lagrange multiplier
to impose the horizontal condition (\ref{horizontal}) on the curve $ V(t) $.
Thus the isoholonomic problem is stated as follows;
find a horizontal curve $ V(t) $ 
that attains an extremal value of the functional (\ref{S})
and satisfies the boundary conditions 
(\ref{close}) and (\ref{hol time evol}).

\section{Derivation and solution of the Euler-Lagrange equation}

We derive the Euler-Lagrange equation associated the functional $ S $
and solve it explicitly.
A variation of the curve $ V(t) $ is defined by
an arbitrary smooth function $ \eta(t) \in \mathfrak{u}(N) $ 
such that $ \eta(0) = \eta(T) = 0 $
and an infinitesimal parameter $ \epsilon \in \R $ as
\begin{equation}
	V_\epsilon (t) = 
	( 1 + \epsilon \eta(t) ) V(t).
\end{equation}
By substituting $ V_\epsilon (t) $ into (\ref{S}) and
differentiating with respect to $ \epsilon $, 
the extremal condition yields
\begin{equation}
	0 = 
	\left. \frac{d S}{d \epsilon} \right|_{\epsilon = 0}
	=
	\int_0^T
	\tr \Big\{
		\dot{\eta} 
		( V \dot{V}^\dagger 
		- \dot{V} V^\dagger 
		- V {\Omega} V^\dagger )
	\Big\} \, dt.
\end{equation}
Thus we obtain the Euler-Lagrange equation
\begin{equation}
		\frac{d}{dt}
		( \dot{V} V^\dagger - V \dot{V}^\dagger 
		+ V {\Omega} V^\dagger )
		= 0.
		\label{extremal}
\end{equation}
The extremal condition with respect to $ {\Omega}(t) $ 
reproduces the horizontal equation
$ V^\dagger  \dot{V} = 0 $.

Next, we solve the equations (\ref{horizontal}) and (\ref{extremal}).
The equation (\ref{extremal}) is integrated to yield
\begin{equation}
		\dot{V} V^\dagger - V \dot{V}^\dagger
		+ V {\Omega} V^\dagger 
		= \mbox{const}
		= X \in \mathfrak{u}(N).
		\label{extremal integrated}
\end{equation}
Conjugation of the horizontal condition (\ref{horizontal}) yields
$ \dot{V}^\dagger V = 0 $.
Then, by multiplying $ V $ on (\ref{extremal integrated}) from the right
we obtain
\begin{equation}
	\dot{V} + V {\Omega} = X V.
	\label{extremal integrated 2}
\end{equation}
By multiplying $ V^\dagger $ on (\ref{extremal integrated 2}) from the left
we obtain
\begin{equation}
	{\Omega} 
	= V^\dagger X V.
	\label{extremal integrated 3}
\end{equation}
We can show $ \dot{{\Omega}} = 0 $
by a straightforward calculation.
Hence, $ {\Omega}(t) $ is actually a constant matrix.
The solution of 
(\ref{extremal integrated 2}) and (\ref{extremal integrated 3}) is
\begin{equation}
	V(t) = e^{t X} \, V_0 \, e^{-t {\Omega}},
	\qquad
	{\Omega} = V_0^\dagger X V_0.
	\label{extremal solution}
\end{equation}
We call this solution the horizontal extremal curve.
Then (\ref{extremal integrated}) becomes
$$
	( XV - V {\Omega} ) V^\dagger
	- V ( - V^\dagger X + {\Omega} V^\dagger )
	+ V {\Omega} V^\dagger 
	= X,
$$
which is arranged as
\begin{equation}
	X -
	( V V^\dagger X 
	+ X V V^\dagger
	- V V^\dagger X V V^\dagger )
	= 0.
	\label{constraint}
\end{equation}
Here we used (\ref{extremal integrated 3}). We may take, without loss of
generality,
\begin{equation}
	V_0 = 
	\left(
		\begin{array}{c}
		I_k \\ 0
		\end{array}
	\right)
	\in S_{N,k} (\C)
	\label{V_0}
\end{equation}
as the initial point.
We can parametrize $ X \in \mathfrak{u}(N) $, 
which satisfies (\ref{extremal integrated 3}), as
\begin{equation}
	X = 
	\left(
		\begin{array}{cc}
		{\Omega}   & W \\
		-W^\dagger & Z 
		\end{array}
	\right)
	\label{X}
\end{equation}
with $ W \in M(k,N-k;\C) $ and $ Z \in \mathfrak{u} (N-k) $.
Then the constraint equation 
(\ref{constraint}) implies that 
$ 
	Z = 0.
$ 
Finally, we obtained a complete set of solution
(\ref{extremal solution})
of the horizontal extremal equation (\ref{horizontal}) and (\ref{extremal}).

\section{Solution to the boundary value problem}
The remaining problem is to find the controller matrices
$ {\Omega} $ and $ W $
that satisfy the closed loop condition
\begin{equation}
	V(T) V^\dagger (T) 
	= e^{TX} V_0 V_0^\dagger e^{-TX} 
	= V_0 V_0^\dagger
	\label{loop condition}
\end{equation}
and the holonomy condition
\begin{equation}
	V^\dagger_0 \, V(T)
	= V^\dagger_0 \, e^{T X} \, V_0 \, e^{-T {\Omega}} 
	= U_{\rm gate}
	\label{holonomy condition}
\end{equation}
for a requested unitary gate $ U_{\rm gate} \in U(k) $.
Montgomery\cite{Montgomery} presented this boundary value problem 
as an open problem.
Here we give a prescription to construct a controller matrix $ X $
that produces the specified unitary gate $ U_{\rm gate} $.
It turns out that the working space should have a dimension
$ N \ge 2k $ to apply our method.
In the following we assume that $ N = 2k $.
The time interval is normalized as $ T = 1 $.

Our method consists of three steps.
In the first step, 
we diagonalize a given unitary matrix $ U_{\rm gate} \in U(k) $ as
\begin{equation}
	R^\dagger U_{\rm gate} R 
	= U_{\rm diag}
	= \mbox{diag} ( e^{i \gamma_1}, \dots, e^{i \gamma_k} )
	\qquad
	( 0 \le \gamma_j < 2\pi )
	\label{det0}
\end{equation}
with $ R \in U(k) $.
The small circle is a circle in a two-sphere $ \C P^1 \subset G_{N,k} (\C) $
that surrounds a solid angle which is equal to twice of the Berry phase.
In the second step, combining $ k $ small circles 
we construct $ k \times k $ matrices
\begin{equation}
	{\Omega}_{\rm diag} 
	= \mbox{diag} ( i \omega_1, \dots, i \omega_k ),
	\qquad
	W_{\rm diag} 
	= \mbox{diag} ( i \tau_1, \dots, i \tau_k ),
\end{equation}
with $ \omega_j = 2 ( \pi - \gamma_j) $ and 
$ \tau_j = e^{i \phi_j} \sqrt{ \pi^2 - (\pi - \gamma_j)^2 } $.
We combine them into a $ 2k \times 2k $ matrix 
$$ 
	X_{\rm diag} = 
	\left( 
		\begin{array}{cc}
		{\Omega}_{\rm diag} & W_{\rm diag}  \\
		- W^\dagger_{\rm diag} & 0
		\end{array}
	\right).
$$ 
In the third step, we construct the controller $ X $ as
\begin{equation}
	X 
	= 
	\left( 
		\begin{array}{cc}
		R & 0 \\
		0 & I_k
		\end{array}
	\right)
	\left( 
		\begin{array}{cc}
		{\Omega}_{\rm diag} & W_{\rm diag}  \\
		- W^\dagger_{\rm diag} & 0
		\end{array}
	\right)
	\left( 
		\begin{array}{cc}
		R^\dagger & 0 \\
		0 & I_k
		\end{array}
	\right)
	= 
	\left( 
		\begin{array}{cc}
		R {\Omega}_{\rm diag} R^\dagger & R W_{\rm diag}  \\
		- W^\dagger_{\rm diag} R^\dagger & 0
		\end{array}
	\right).
	\label{det3}
\end{equation}
In the paper\cite{isoholo} we calculated explicitly controllers
of various unitary gates; 
the controlled NOT gate, the discrete Fourier transformation gate and so on.

\section{Conclusion}
We formulated and solved the isoholonomic problem in the homogeneous fiber bundle.
The problem was reduced to a boundary value problem of the horizontal extremal equation.
We determined the control parameters that satisfy the boundary conditions.
This result is applicable for producing arbitrary unitary gates.

\end{document}